\documentclass[aps]{revtex}
\usepackage{graphicx,ams}

\begin{document}

\renewcommand{\R}{\mbox{$\cal R$}}
\newcommand{\Mp}{\mbox{$M_{\rm Pl}$}}
\newcommand{\sign}{\mbox{sign}}
\newcommand{\erfcx}{\mbox{erfcx}}
\newcommand{\erfc}{\mbox{erfc}}
\newcommand{\hypergeom}{\mbox{F}}
\newcommand{\VphiO}{\mbox{$\left.V_{,\phi}\right|_{t=0}$}}
\newcommand{\Ceven}{\mbox{$C_{\rm even}$}}
\newcommand{\Codd}{\mbox{$C_{\rm odd}$}}
\newcommand{\Cgrowbef}{\mbox{$C^-_{\rm grow}$}}
\newcommand{\Cdecaybef}{\mbox{$C^-_{\rm decay}$}}
\newcommand{\Cgrowaft}{\mbox{$C^+_{\rm grow}$}}
\newcommand{\Cdecayaft}{\mbox{$C^+_{\rm decay}$}}

\newcommand{\Phieven}{\mbox{$\Phi_{\rm even}$}}
\newcommand{\Phiodd}{\mbox{$\Phi_{\rm odd}$}}

\newcommand\eq[1]{Eq.~(\ref{#1})}
\newcommand\eqs[2]{Eqs.~(\ref{#1}) and (\ref{#2})}
\newcommand\eqss[3]{Eqs.~(\ref{#1}), (\ref{#2}) and (\ref{#3})}
\newcommand\eqsss[4]{Eqs.~(\ref{#1}), (\ref{#2}), (\ref{#3})
and (\ref{#4})}
\newcommand\eqssss[5]{Eqs.~(\ref{#1}), (\ref{#2}), (\ref{#3}),
(\ref{#4}) and (\ref{#5})}
\newcommand\eqsssss[6]{Eqs.~(\ref{#1}), (\ref{#2}), (\ref{#3}),
(\ref{#4}), (\ref{#5}) and (\ref{#6})}
\newcommand\eqst[2]{Eqs.~(\ref{#1})--(\ref{#2})}
\newcommand\<{\left[}
\renewcommand\>{\right]}

\newcommand\slabel[1]{\label{#1}}

\title{Cosmological Perturbations Through a General Relativistic
Bounce}

\author{Christopher Gordon and Neil Turok}

\address{DAMTP, Centre for Mathematical Sciences, Cambridge
University, Cambridge, CB3 0WA, United Kingdom}

\maketitle

\begin{abstract}
The ekpyrotic and cyclic universe scenarios have revived the idea
that the density perturbations apparent in today's universe could have been 
generated in a `pre-singularity' epoch before the big bang.
These scenarios provide explicit mechanisms whereby a scale
invariant spectrum of adiabatic perturbations 
may be
generated 
without the need for cosmic inflation,
albeit in a phase preceding the hot big bang singularity.
A key question they
face 
is whether there exists a unique prescription for following
perturbations through the bounce, an 
issue which is not yet definitively
settled. This goal of this paper is more modest, namely to study
a bouncing Universe model in which
neither General Relativity nor the Weak Energy Condition is violated.
We
show that a perturbation which is pure growing mode before the bounce
does not match to a pure decaying mode perturbation after the
bounce. Analytical estimates of when the comoving curvature
perturbation varies around the bounce are given. It is found that
in general 
it is necessary to evaluate the evolution of the perturbation through
the bounce in detail rather than using matching conditions.
\end{abstract}

\section{Introduction}

In the inflationary Universe scenario, large scale
structure is generated by the stretching of subatomic scale
quantum fluctuations to macroscopic length scales, during
an epoch of superluminal expansion supposed to
have occurred prior to the radiation dominated era. 
Indeed such expansion appears to be essential to the creation of
correlations on super-Hubble radius scales, 
a feat prohibited by causality in the standard hot big bang era.

However, cosmic inflation does not resolve the problem of the initial
singularity, and it seems clear that a more complete theory than
inflation is needed to deal with it. But if time existed before the
the initial singularity, it is a logical possibility that the large
scale structure and density perturbations we see today were generated
during this pre-singularity epoch. The ekpyrotic and cyclic
scenarios\cite{ekpyrotic,cyclic} provide explicit realisations of this
idea (pre-figured in the `pre-big bang' scenario of Veneziano et
al.\cite{pbb}). In the ekpyrotic and cyclic scenarios, scale invariant
large scale density perturbations are generated during a phase which
is contracting (in the Einstein frame) before the bounce to
expansion\cite{ekpyrotic}. The task of matching these perturbations
across the bounce to the expanding epoch is highly challenging, since
general relativity must break down there, and the use of string theory
methods will ultimately be essential. There are indications that for
the particular type of singularity involved here (namely the collapse
of a single extra dimension) the divergences are relatively
weak\cite{ekperts}, but the matching issue remains unsettled at the
present time\cite{ekperts2,MartinPeter,Lyth}.

In a contracting universe, the growing mode adiabatic density
perturbation corresponds to a shift in the time to the
`big crunch'. A perturbation in the space curvature, however, 
is a decaying mode perturbation because it becomes increasingly
irrelevant as the scale factor shrinks. (These statements both hold in 
Einstein frame and comoving gauge).  However, in an expanding universe, the situation
is reversed. A 
time delay is now a decaying mode perturbation and the curvature perturbation
is the growing mode. This has led several authors to suggest
that in the ekpyrotic and cyclic scenarios, where
scale invariant density perturbations are generated in the
collapsing phase growing mode, that these perturbations would match
on to pure decaying mode perturbations in the subsequent 
expanding phase. Indeed this result is obtained if one insists on matching
the curvature perturbation across the bounce \cite{ekperts2}.

However, this behaviour would appear surprising from a more physical
viewpoint. In the models being considered, gravity is attractive
throughout. The growing mode instability is driven by 
gravitational attraction and it is physically implausible that it should
precisely
reverse at a bounce so as to dissipate after the bounce and
render the final universe 
perfectly homogeneous. 

In this paper we address this issue by considering a model in
which the issue can be definitively settled within conventional
general relativity. 
It is well known that a massive scalar field in a closed
Friedman-Robertson-Walker (FRW) Universe can lead to a bounce without
a singularity, and without 
violating the Weak Energy Condition. This model
has a long history starting with Schr\"{o}dinger
\cite{Schrodinger}, however he did not include the back-reaction of
the scalar field on the background scale factor.  It was also employed in
semi-classical studies of quantum effects
(see
for example \cite{Fulling}), and of course is a common model in
quantum cosmology and studies of the no boundary proposal
(see for example \cite{Hawking}, and more recently \cite{Gratton}). 
The classical dynamics for the spatially homogeneous case
have also been investigated quite extensively, see for example
\cite{bgrefs}.

A preliminary study of cosmological perturbations in
such a Universe (and in other bouncing models) was performed 
 by Hwang and Noh
\cite{Hwang}, with inconclusive results since 
the usual large scale approximations
break down around the bounce. We develop new approximations that
do not break down around the bounce, and using numerical simulations,
accurately determine the propagation of linear perturbations
through the bounce. We show that growing mode perturbations developed in the
collapsing phase do not match to pure decaying mode perturbations in the
expanding phase, thus there is no contradiction with the 
physical argument given above. Of course, this does not at all
prove that the same is true in the ekpyrotic/cyclic models: as the original papers
made clear, an unambiguous matching condition is still required. 
But the present work does show that propagation of a growing
mode perturbation across a bounce is possible at least in 
this toy model.

We would also like to briefly mention recent work on the ekpyrotic/cyclic
scenarios. 
Durrer and Vernizzi \cite{Durrer} showed how matching across the
bounce depends on what surface the matching is done on and the intrinsic 
stress energy of that surface. They showed that 
only for the special case
of matching on constant energy density surfaces with no surface
tension that the growing mode in the collapsing phase matches
completely onto the decaying mode in the expanding phase. Generally,
some of the growing mode in the collapsing phase will go to the
growing mode in the expanding phase. We should also mention a recent 
study of the 
possibility of isocurvature perturbations in the Ekpyrotic model
\cite{Riotto}.
Peter and Pinto-Neto \cite{Peter1} studied the matching problem in
a hydro-dynamical
fluid, and in a second publication employed a scalar field with 
negative kinetic energy in order to obtain a regular bounce
\cite{Peter3}. The problem of matching
quantum fields across a bouncing universe of the type relevant
to the ekpyrotic/cyclic scenarios has been considered 
by Tolley and Turok \cite{Tolley}, and a number of studies
have been made of analogous bouncing models within string theory
\cite{stringmodels}.

\section{Model}

The equation for the scale factor in a FRW Universe is given by
\begin{equation}
\ddot{a} = -\frac{1}{6\Mp^2}(\rho + 3p)a
\slabel{rachaduri}
\end{equation}
where $a$ is the scale factor, dot denotes a derivative with respect
to time, $\rho$ is the density, $p$ is the pressure and
$\Mp=1/\sqrt{8\pi G}$ is the reduced Planck mass. As the scale factor is
always positive and $\dot{a}$ is negative for a collapsing Universe,
it follows from \eq{rachaduri} that the Strong Energy Condition
($\rho + 3p \ge 0$) must be violated  for $\ddot{a}>0$. Which is what is
needed for a bounce at a finite scale factor to be possible. This is
the same as the condition for inflation. It is satisfied by a scalar
field that is potential dominated as 
\begin{equation}
\rho = \frac{1}{2}\dot{\phi}^2 + V
\slabel{density}
\end{equation}
and
\begin{equation}
p = \frac{1}{2}\dot{\phi}^2 - V
\slabel{pressure}
\end{equation}
where $\phi$ is the value of the scalar field and V is its
potential. For simplicity we shall only consider  a quadratic potential
\begin{equation}
V = \frac{1}{2} m^2 \phi^2
\slabel{potential}
\end{equation}
where $m$ is the constant mass of the scalar field.

Using the Friedman equation
\begin{equation}
H^2 = \frac{1}{3\Mp^2} \rho - \frac{K}{a^2}
\slabel{Friedman}
\end{equation}
it can be seen that the background curvature ($K$) must be positive for
the Universe to bounce. This is because at the bounce $H=0$ and
$\rho>0$. As $\rho + p \ge 0$, it follows that the Weak Energy
Condition is never violated by this model.

Although the above arguments show that it is possible for a closed
Universe with a scalar field to bounce, it does depend on the initial
conditions. The scalar field evolution is given by the Klein-Gordon
equation
\begin{equation}
\ddot{\phi} + 3H\dot{\phi} + V_{,\phi} = 0
\slabel{KG}
\end{equation}
where $V_{,\phi} = dV/d\phi$. In a expanding Universe $H>0$ and so the
$3H\dot{\phi}$ acts as a friction term. While in a collapsing Universe
$H<0$ and so the $3H\dot{\phi}$ acts as an anti-friction term.

So, it is possible that the the system will always be kinetic energy
dominated and go to $\phi = \pm \infty$ and $a=0$ 
instead of bouncing. 
It is only for a small subset of initial conditions that
such a Universe would bounce \cite{bgrefs}. 
The bounce will be symmetric in the background quantities if
$\dot{\phi}=0$ when $\dot{a}=0$. We use this case to study the
evolution of perturbations through the bounce which is taken to occur
at $t=0$.
Figure \ref{sketch}
summarises the sequence of events.
\begin{figure}
\begin{center}
\includegraphics[width=8cm]{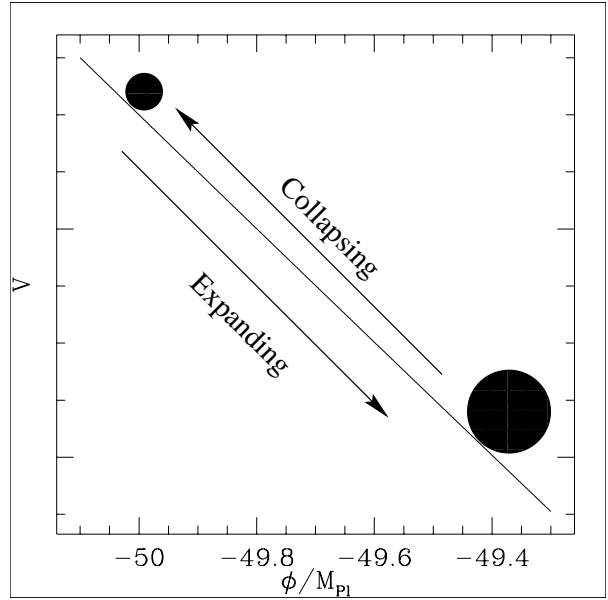}
\end{center}
\caption{\slabel{sketch} A sketch of the background dynamics of the
model investigated in this paper. The scalar field is represented by
the ball and $\ln(a)$ is proportional to the size of the ball. The
scale factor changed by approximately 15 e-folds from its minimum to
maximum size in the 
numerical integration.
First the scalar field rolls up the potential as the
Universe collapses. Then when the Universe reaches a non-zero scale
factor $a_0$, it starts to expand again and roll down the
potential. The background dynamics are time-symmetric around the
bounce.}
\end{figure}

Around the bounce we can make the approximation $\dot{\phi}\approx 0$
and $\phi=\phi_0$ where subscript $0$ is taken to the value of the
quantity at the bounce $(t=0)$. Using
\eqss{rachaduri}{density}{pressure} this gives
\begin{equation}
\slabel{abounce}
a \approx a_0 \cosh(t/a_0).
\end{equation}
This is the same as the equation for a closed de Sitter space. If we
choose $K=1$ then from \eqs{potential}{Friedman}
\begin{equation}
a_0 = \Mp \sqrt{\frac{3}{V_0}} =\frac{\sqrt{6}}{m}  \frac{\Mp}{|\phi_0|} 
\label{a0}
\end{equation}
The Hubble parameter around the bounce is then given by
\begin{equation}
\slabel{Hbounce}
H = \frac{1}{a_0} \tanh(t/a_0).
\end{equation}
So the time scale around the bounce is $a_0$.
In order to approximate the evolution of the scalar field around the
bounce we rewrite \eq{KG} as 
\begin{equation}
(\dot{\phi} a^3\dot{)} = -V_{,\phi} a^3.
\end{equation}
Integrating both sides gives
\begin{equation}
\dot{\phi} a^3 = -\int_0^t V_{,\phi} a^3 \, dt
\end{equation}
where $(\dot{\phi})_0=0$ was used. Using $V_{,\phi}\approx
(V_{,\phi})_0$ and \eqs{potential}{abounce} we get
\begin{equation}
\slabel{phibounce}
\phi \approx \frac{1}{3} a_0^2 m^2\phi_0 \left( \frac{1}{\cosh^2(t/a_0)} -
\ln(\cosh(t/a0)) -1\right)  + \phi_0.
\end{equation}
Away from the bounce ($t\gtrsim a_0$), this equation can be approximated by
\begin{equation}
\phi \approx \phi_0\left(2\frac{\Mp^2}{\phi_0^2}\left(-1+\ln(2)-\frac{|t|}{a_0}\right)+1\right).
\slabel{phisr}
\end{equation}
This is the same equation that is obtained by using the slow roll
approximations to \eqs{Friedman}{KG} (see for example \cite{LL})
\begin{equation}
3H\dot{\phi} + V_{,\phi} \approx 0
\end{equation}
and
\begin{equation}
\slabel{Hsr}
H^2 \approx \frac{1}{3\Mp^2} V.
\end{equation}
Which will be a good approximation when the slow roll parameters \cite{LL}
\begin{equation}
\epsilon = \frac{\Mp^2}{2} \left( \frac{V_{,\phi}}{V} \right)^2
\end{equation}
and 
\begin{equation}
|\eta| = \Mp^2 \frac{1}{V} \left| \frac{d^2V}{d\phi^2}\right|
\end{equation}
are small and there has been enough expansion to make the $K/a^2$ in
\eq{Friedman} negligible. For the quadratic potential given in
\eq{potential} both conditions become
\begin{equation}
\slabel{slowrollcond}
\phi^2 \gtrsim 2\Mp^2.
\end{equation}
For $|t|\gtrsim a_0$, 
a good approximation for $H$ is obtained by substituting
\eq{phisr} into \eq{Hsr}
\begin{equation}
H(|t| \gtrsim a_0) \approx \frac{\sign(t)}{a_0}\left|
2\left(\frac{\Mp}{\phi_0}\right)^2\left(-1-
\frac{|t|}{a_0}+ \ln(2)\right) + 1 \right|
\slabel{Hsr1}
\end{equation}
This can then be integrated to obtain an expression for $a(t)$ and the
integration constant can be found by matching to the bounce
approximation \eq{abounce}
\begin{equation}
a(|t|\gtrsim a_0) \approx \frac{a_0}{2} \exp\left(\frac{|t|}{a_0}\left(\ln(4)-2- \frac{|t|}{a_0} \right) \frac{\Mp^2}{\phi_0^2}+
\frac{|t|}{a_0} \right).
\slabel{asr}
\end{equation}
If we further assume $|\phi_0|/\Mp \gg |t|/a_0$ then
\eqss{phisr}{Hsr1}{asr} can be simplified to 
\begin{equation}
\phi(|t|\gtrsim a_0) \approx \phi_0, \quad H(|t|\gtrsim a_0)
\approx \frac{\sign(t)}{a_0}, \quad
\left.\frac{dH}{dt}\right|_{|t|\gtrsim a_0} \approx -2 \frac{\Mp^2}{\phi_0^2} \frac{1}{a_0^2}
, 
\quad a(|t|\gtrsim a_0) \approx \frac{a_0}{2}
\exp\left(\frac{|t|}{a_0}\right). \slabel{sr}
\end{equation}
In which case $a_0$ is also the time scale away from the
bounce.

\section{Perturbation Equations}
For studying the evolution of the perturbations we use the Bardeen
formulism \cite{Bardeen,MFB}. For purely scalar perturbations in the
zero-shear (conformal Newtonian) gauge, the perturbed metric is given by
\begin{equation}
\slabel{metric}
ds^2 = -(1+2\Phi)dt^2 + a^2(t)(1-2\Psi)\gamma_{ij} dx^i dx^j
\end{equation}
where $x^i$ is the comoving spatial coordinate and for a closed
Universe 
\begin{equation}
\gamma_{ij} = \delta_{ij}\left(1 + \frac{1}{4} K x^k x_k\right)^{-2}.
\end{equation}
where Latin indices range from 1 to 3, $\Phi$ and $\Psi$ are the
metric perturbations and $\delta_{ij}$ is the Kronecker delta.  For a
scalar field there is no anisotropic stress and so there is the
further simplification $\Psi=\Phi$ \cite{MFB}.

The equation for the evolution of $\Phi$ can be derived from the
Einstein equations \cite{MFB,BGT} and is
\begin{equation}
\slabel{Phi}
      \ddot{\Phi} + \left(H - 2\frac{
     \ddot{\phi}}{\dot{\phi}}\right)\dot{\Phi} + \frac{1}{a^2}(-\nabla^2
     -4K)\Phi + 2\left(\dot{H} -H\frac{   \ddot{\phi}}{\dot{\phi}}
     \right)\Phi = 0
\end{equation}
The usual harmonic decomposition is performed on the perturbations so
that in a closed Universe the eigenvalues of $-\nabla^2$ are $(k^2 =
[n^2 - 1]K)$ \cite{abbotschafer} where $n\ge 3$ is an integer. The
$n=1$ mode is homogenous and the $n=2$ mode is a pure gauge mode
\cite{Bardeen}. We will be looking at the evolution of an individual
mode with time. So the dependence of $\Phi$ will be in terms of $n$
and $t$.

An approximation for $\Phi$ around the bounce ($|t| \lesssim a_0$) can be found by
substituting \eqs{phibounce}{abounce} and $K=1$  into \eq{Phi} and
working with the harmonic components of $\Phi$
\begin{eqnarray}
\slabel{Phiode1}
&&\ddot{\Phi} +
\left(\frac{\tanh(t/a_0)}{a_0}-2\frac{\ddot{\phi}}{\dot{\phi}} \right)
\dot{\Phi}+\frac{n^2-5}{a_0^2\cosh^2(t/a_0)}\Phi \\ \nonumber
&&\quad\quad\quad+2\left(\frac{1}{a_0^2}-\frac{\tanh^2(t/a_0)}{a_0^2}-\frac{\tanh(t/a_0)}{a_0}
\frac{\ddot{\phi}}{\dot{\phi}}\right)\Phi = 0
\end{eqnarray}
The equation can
be reduced to a solvable form by a non-linear transform of the time
coordinate
\begin{equation}
\slabel{nu}
\nu=\ln\left(\tanh\left(\frac{1}{2}\frac{t}{a_0} - \frac{\pi}{4}i \right) \right)
\end{equation}
where $i=\sqrt{-1}$. Then \eq{Phiode1} becomes
\begin{equation}
\frac{\partial^2\Phi}{\partial \nu^2} + 
\frac{6\left( -1+e^{2\nu} \right)
\left( 1 + e^{4\nu}  \right)}
{\left(-1 + 4e^{2\nu} - e^{4\nu} \right)\left( 1 + e^{2\nu} \right)}
\frac{\partial\Phi}{\partial \nu} +
\left( -n^2 + 5 - 
\frac{4(1+e^{2\nu})^2}
{-1 + 4e^{2\nu} - e^{4\nu}}
 \right) \Phi  = 0.
\end{equation}
Which has the solution
\begin{equation}
\slabel{Phibounce}
\Phi(|t|\lesssim a_0) \approx C_1\Phi_1 + C_2 \Phi_1^*
\end{equation}
where $C_1$ and $C_2$ are integration constants, superscript $*$  denotes the
complex conjugate and 
\begin{equation}
\slabel{Phi1}
\begin{array}{l}
\Phi_1 = e^{(n-3)\nu } 
\left[   
-(n-1)(n-2)e^{6\nu} 
+ 3(n+1)(n-2)e^{4\nu} \right. \\
\quad + \left. 3(n+2)(n-1)e^{2\nu} 
- (n+2)(n+1)
 \right].
\end{array}
\end{equation}
We will use this solution to check our numerical calculations in
Section~\ref{numerics}.

For $|t|\gtrsim a_0$ the Universe is effectively flat ($K=0$) and we
can use the the background approximations given in
\eqss{phisr}{Hsr1}{asr}. As seen from \eq{Phi}, we can also treat the
perturbations as if we were in a flat Universe provided $n^2-1 \gg 4$.
In which case the usual transformation can be used \cite{MFB,MartinSchwarz}
\begin{equation}
 u = \Phi / \dot{\phi}
\slabel{u}
\end{equation}
which has the equation 
\begin{equation}
u'' + (k^2 - \theta''/\theta)u=0
\slabel{odeu}
\end{equation}
where prime indicates the derivative with respect to conformal
time $\tau$ with $a d\tau = dt$ and 
\begin{equation}
\theta = \frac{H}{a\dot{\phi}}\,.
\slabel{theta}
\end{equation}
The `large scale
condition' is    
\begin{equation}
k^2 \ll \theta''/\theta.
\slabel{lscond}
\end{equation}
A simple expression for the time spans that the large scale condition
is satisfied can be obtained by 
assuming that $|\phi_0|/\Mp \gg |t|/a_0$ and using
\eqss{sr}{theta}{lscond}
which gives
\begin{equation}
\frac{|t|}{a0} \gtrsim  \ln \left(k
\frac{|\phi_0|}{\Mp}\right).
\slabel{lsscond1}
\end{equation}
As is expected the larger $k$ the further from the bounce the large
scale approximation becomes valid. Also, the large scale condition is
always violated for $|t|>a_0$ as $k^2>= 3^2-1$ and $|\phi_0|>\Mp$. When
the large scale condition is not satisfied the solutions of \eq{u} and
therefore \eq{Phi} are clearly oscillatory with frequency $k$.

If the large scale condition is satisfied then we can approximate
the solution to \eq{odeu} by a truncated series of $k^2$:
\begin{equation}
u = \sum_{j=0}^\infty C_{2j}(\tau) k^{2j}. 
\slabel{kseries}
\end{equation}
Substituting this into \eq{odeu} and matching coefficients on the left
and right side gives the zeroth order approximation
\begin{equation}
u = C_3 \theta + C_4 \theta \int \frac{1}{\theta^2} \, d\tau.  
\slabel{usoln}
\end{equation}
Using \eqsssss{phisr}{Hsr1}{asr}{u}{theta}{usoln} and redefining the constants $C_3$
and $C_4$ gives
\begin{equation}
\slabel{Phisr}
\Phi(|t|\gtrsim a_0) \approx C_3 \frac{H}{a}
+ C_4 \left(1 - \sqrt{\pi} \psi \erfcx(\psi)  \right)
\end{equation}
where 
\begin{equation}
\psi = \frac{1}{2} \frac{|\phi|}{\Mp}
\end{equation}
and
\begin{equation}
\erfcx(\psi) = e^{\psi^2} \erfc(\psi)
\end{equation}
is the scaled complementary error function. 
If we further assume $|\phi_0|/\Mp \gg |t|/a_0$, then substituting
\eq{sr} into \eq{Phisr} gives \cite{AS}
\begin{equation}
\Phi(|t|/a_0 \gtrsim 1) \approx 2\left(C_3 \sign(t) \frac{1}{a_0^2}
\exp\left( -|t|/a_0 \right) + C_4 \frac{\Mp^2}{\phi_0^2} \right). 
\slabel{Phisr1}
\end{equation}
Although the form of the solution (\eq{Phisr}) is the same before and
after the bounce, it is not necessary that the integration constants
($C_3$ and $C_4$) should be the same. The slow-roll approximation for
\eq{Phi} is not the same before and after the bounce due to the
presence of the absolute value signs in \eqs{phisr}{Hsr1}.
So we will have four integration constants:
\begin{equation}
C_3 = \left\{
\begin{array}{ll}
\Cgrowbef, & t \lesssim -a_0   \\
\Cdecayaft, & t \gtrsim a_0
\end{array}
\right.
\slabel{C3}
\end{equation}
and
\begin{equation}
C_4 = \left\{
\begin{array}{ll}
\Cdecaybef, & t \lesssim -a_0   \\
\Cgrowaft, & t \gtrsim a_0.
\end{array}
\right.
\slabel{C4}
\end{equation}
The $+$ and $-$ superscript indicate after and before the bounce. The
`grow' and `decay' subscripts indicate that the constant is 
the
amplitude of the growing and decaying mode respectively.

In principle there should be two constraint equations relating the
four integration constants as \eq{Phi} is second order and so should
only have two integration constants. That is, there should be a
matching rule expressing $\Cdecayaft$ and $\Cgrowaft$ in terms of
$\Cdecaybef$ and $\Cgrowbef$.


\section{Numerical solutions}
\label{numerics}
Although \eqs{Phibounce}{Phisr} provide good approximations around and
away from the bounce. They do not overlap sufficiently to provide a
solution for the whole domain of interest for all initial
conditions. 
In principle, the stage before the global curvature becomes important,
$|t|\gtrsim a_0$, and when the large scale condition is violated,
\eq{lsscond1}, could also be included in the matching. But the
analytical approximations are found still not to be sufficiently
accurate in the overlap regions for general initial conditions.
Therefore it is also necessary to solve \eq{Phi}
numerically.

Before doing a numerical integration it is important to identify if
there are any singularities in the coefficients of \eq{Phi}.
A series solution for \eqs{rachaduri}{KG} is
\begin{equation}
\slabel{phiseries}
\phi = \phi_0 \left[
1 - 3\frac{\Mp^2}{\phi_0^2}\left(\frac{t}{a_0}\right)^2 +O(4)
\right]
\end{equation}
and
\begin{equation}
\slabel{aseries}
a = a_0 \left[
1+\frac{1}{2} \left(\frac{t}{a_0}\right)^2 + O(4)
\right].
\end{equation}
Substituting \eqs{phiseries}{aseries} and $\tau=t/a_0$ into
\eq{Phi} gives
\begin{equation}
\frac{\partial^2 \Phi}{\partial \tau^2} 
- \frac{4}{(2+\tau^2)\tau} \frac{\partial \Phi}{\partial \tau}
- 4\frac{-n^2+5+2\tau^2}{(2+\tau^2)^2} \Phi =0.
\slabel{Phiodeseries}
\end{equation}
As can be seen there is a singularity in the coefficient for
$\partial\Phi /\partial \tau$ at $\tau=0$. This means that it is not
possible to numerically integrate \eq{Phi} through the bounce.
However, we can solve \eq{Phiodeseries} to give an
analytical solution very close to the bounce
\begin{equation}
\slabel{Phiodeseriessoln}
\begin{array}{l}
\Phi = (2+\tau^2)^{-N} \\ \quad \left\{
C_5
\hypergeom\left(-N+\frac{1}{4}(-1
+ \sqrt{33}),
-N-\frac{1}{4}(1+\sqrt{33}),-\frac{1}{2},-\frac{1}{2}\tau^2\right)
 \right. \\ \quad + \left.
C_6 \tau^3 \hypergeom\left(-N+\frac{1}{4}(5-\sqrt{33}), -N+\frac{1}{4}(5+\sqrt{33}),\frac{5}{2},-\frac{1}{2}\tau^2\right)
\right\}
\end{array}
\end{equation}
where $\hypergeom(\cdot)$ is the hyper-geometric function, $N =
\sqrt{(n^2-1)/2}$ and $C_5$ and $C_6$ are integration constants. A
series approximation can be used to evaluate \eq{Phiodeseriessoln}
around the origin
\begin{equation}
\slabel{Phiseries}
\Phi = \Ceven \Phieven + \Codd \Phiodd
\end{equation}
where $\Ceven$ and $\Codd$ are integration constants and
\begin{equation}
\slabel{Phieven}
\Phieven = \frac{1}{2}(n^2-5) \left( \frac{t}{a_0} \right)^2 + 1 + O(4)
\end{equation}
and
\begin{equation}
\slabel{Phiodd}
\Phiodd = \left( \frac{t}{a_0} \right)^3 + O(5).
\end{equation}
\eqss{Phiseries}{Phieven}{Phiodd} can then be used to generate initial
conditions on either side of the bounce for an adaptive step-length
Runge-Kutta numerical integration routine \cite{NR}. We set these
initial conditions for the numerical simulation at $t/a_0 = \pm
10^{-4}$.  The relative accuracy for the Runge-Kutta routine was set
to $10^{-14}$.  The numerical results were found to be insensitive to
the precise values used.

The parameters used for the simulations were $\phi_0 = -50 \Mp$ and
$m=10^{-6}\Mp$.  As the system of equations \eqss{rachaduri}{KG}{Phi}
only depends on $m$ through the combination $mt$, changing $m$ will
just change the scaling of time. As can be seen from
\eqss{nu}{Phibounce}{Phi1}, around the bounce the only dependence of
$\phi_0$ is through the dependence on $a_0$, see \eq{a0}. So around
the bounce changing $\phi_0$ will just change the time scale of the
problem. From \eqs{sr}{Phisr1}, it can be seen that
increasing $\phi_0$ will  effect the time scale through its
effect on $a_0$ (\eq{a0}) and the normalisation of the $C_3$ and $C_4$ modes.

The only other parameters that need to be specified are the initial
values for $\Phi$ and $\dot{\Phi}$. The numerical solution for these
can be found by choosing $\Ceven$ and $\Codd$ in \eq{Phiseries} so
that the numerical extrapolation has the required values of $\Phi$ and
$\dot{\Phi}$ at the initial time. This technique is known as {\em
shooting} \cite{NR}.

Figure \ref{fig:Phi} shows a plot of $\Phi$ for the wave-number
corresponding to $n=3$.
 The initial conditions at $t/a_0=-20$ were set to be a pure growing
mode $\Phi=H/a$ using the shooting method. The decaying mode in the
collapsing phase and the growing mode in the expanding phase was
evaluated by matching \eq{Phisr} at the initial and final times. They
are plotted for those times for which the large scale condition,
\eq{lsscond1}, is satisfied. This corresponds to $|t|/a_0 \gtrsim
5$. Due to the finite accuracy of the shooting algorithm, there is
still a finite amplitude of decaying mode a $t/a_0=-20$. However, by
the time the large scale condition, \eq{lsscond1}, is violated, the
decaying mode is negligible relative to the growing mode.
\begin{figure}
\begin{center}
\includegraphics[width=8cm]{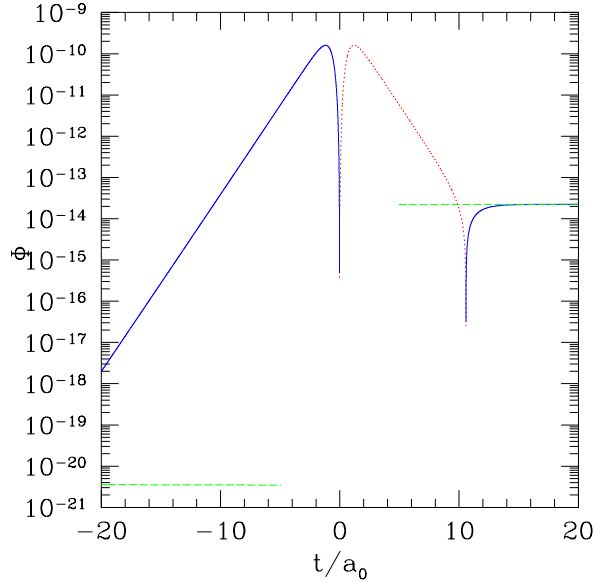}
\end{center}
\caption{\slabel{fig:Phi} A plot  of the absolute
value of the zero
shear curvature perturbation $\Phi$ for the wave number corresponding
to $n=3$. The solid lines are for positive $\Phi$ and the dotted line
is for where $\Phi$ is negative. The collapsing phase decaying mode
and expanding phase growing mode (\eq{Phisr}) are plotted as dashed
lines.  }
\end{figure}

As can be seen from \eqss{Phiseries}{Phieven}{Phiodd}, $\Phi$ and its
first and second time derivatives are finite at $t=0$. This shows that
there are initial conditions that lead to linear perturbations around
the bounce \cite{Lyth,Peter1}. 

\begin{figure}
\begin{center}
\includegraphics[width=8cm]{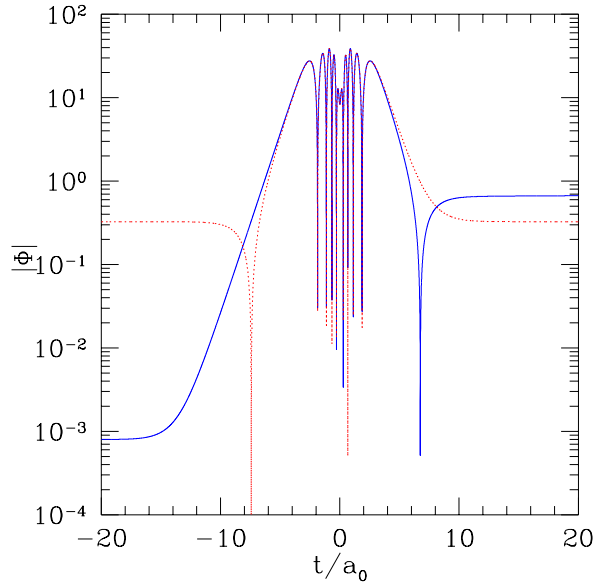}
\end{center}
\caption{\label{n10} A plot of $\Phi$ (solid line) and its
 approximation using \eq{Phibounce} (dotted line) for $n=10$.  }
\end{figure}

Figures \ref{n3} and \ref{n10} show examples of numerical solutions
and the bounce approximation, \eq{Phibounce}, for small and large wave
numbers. The initial conditions for Figure \ref{n3} are the same as
those for Figure \ref{fig:Phi}. In Figure \ref{n10}, the initial conditions
are set so that at $t/a_0=-20$ there is a pure decaying mode,
\eq{Phisr1}, $\Phi = \Mp^2/\phi_0^2$. It is an important confirmation 
of the numerical integrations that they match the bounce approximations
so well for $|t|/a_0 \lesssim 1$. 
\begin{figure}
\begin{center}
\includegraphics[width=8cm]{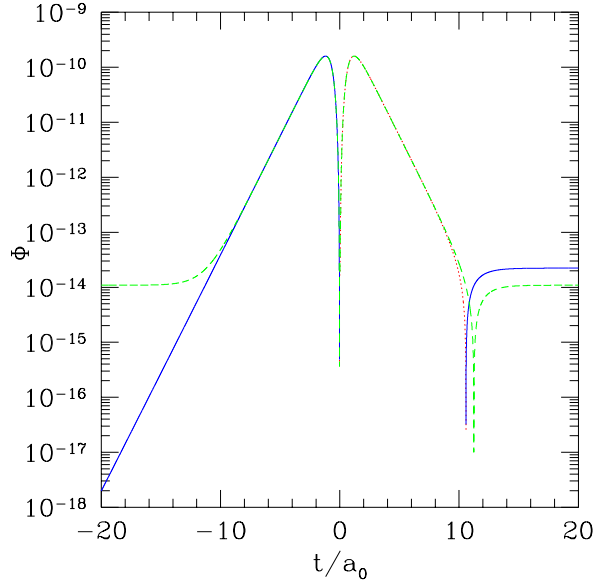}
\end{center}
\caption{\slabel{n3} A plot of the absolute value of the zero shear
curvature perturbation $(\Phi)$ and its approximation around the
bounce, \eq{Phibounce}, for $n=3$. The solid lines are for positive
$\Phi$ and the dotted line is for where $\Phi$ is negative. The
absolute value of the approximation is plotted as a dashed line.}
\end{figure}

\section{Comoving Curvature}
\label{comov}
In inflation, the use of the curvature on constant density
hyper-surfaces ($\zeta_{\rm BST}$) \cite{BST} or the curvature on comoving
hyper-surfaces ($\R$) \cite{Lyth85} is very useful  because
in an expanding flat Universe with no decaying mode or entropy
perturbation they are constant on scales larger than the Hubble
horizon ($1/H$). We will be examining whether this can be used for the
present model.

The comoving curvature perturbation for any background curvature ($K$)
is \cite{Karim}
\begin{equation}
\slabel{R}
\R = \Psi + 2\Mp^2 H \frac{H\Phi + \dot{\Psi}}{\rho + p}
\end{equation}
where for a scalar field $\Psi=\Phi$.
Note that the formula for the comoving curvature giving in Mukhanov
et.~al.\ \cite{MFB} ($\zeta_{\rm MFB}$) only applies for the flat
case ($K=0$).
Alternatively the curvature on constant density hyper-surfaces
\cite{BST} could be used. However, away from the bounce, when the
decaying mode is sub-dominant, they are all approximately equal up to
a change in sign and so they will all serve the same purpose below.

Using \eqssss{phiseries}{aseries}{Phiseries}{Phieven}{Phiodd} in \eq{R} gives
\begin{equation}
\R(0) = \Ceven (2n^2-7)
\end{equation}
which shows that $\R$ is finite at the bounce.

Using \eqsssss{density}{pressure}{potential}{phisr}{Hsr1}{Phisr} into \eq{R} gives
\begin{equation}
\R \approx C_4.
\slabel{Rconst}
\end{equation}
From \eq{Rconst} it appears that $\dot{\R}=0$ when the large scale
condition, \eq{lscond}, holds.
 However, \eq{Rconst} was derived using the zeroth order
truncation of $k^2$, \eqss{kseries}{usoln}{Phisr}. As is well known
\cite{MFB,MartinSchwarz,LL}, $\R$ will not be constant if the series
expansion, \eq{kseries}, is extended to order $k^2$. Alternatively, the
time change of $R$ can be found using the general equation \cite{LL}
\begin{equation}
\dot{\R} = H \frac{\delta p_{\rm comov}}{\rho + p}
\end{equation}
where $\delta p_{\rm comov}$ is the pressure perturbation on comoving
hyper-surfaces. For a single scalar field this becomes \cite{GWBM}
\begin{equation}
\dot{\R} = \frac{H}{\dot{H}} \frac{k^2}{a^2} \Phi.
\slabel{Rdot}
\end{equation}
As $a$ shrinks exponentially as the bounce is approached, \eq{asr}, $\R$
may no longer be approximately constant. Thus, $\R$ can vary on large
scales as the pressure perturbation on comoving hyper-surfaces grows
exponentially as the bounce is approached. 

The pressure perturbation can be decomposed as \cite{LL}
\begin{equation}
\delta p = \frac{\dot{p}}{\dot{\rho}} \delta \rho + \delta p_{\rm nad}
\end{equation} 
where $\delta \rho$ is the density perturbation and $\delta p_{\rm
nad}$ is the gauge invariant non-adiabatic component of the pressure
perturbation. For the quadratic potential in this paper, both the
adiabatic and non-adiabatic components of the pressure perturbation on
comoving hyper-surfaces grow exponentially as the bounce is
approached. Where as for a massless field, such as in the stage just
before the bounce in the Ekpyrotic scenario \cite{ekpyrotic} there is
no non-adiabatic component to the pressure perturbation \cite{GWBM}
and it is only the adiabatic part of the pressure perturbation that
grows and causes $\R$ to vary \cite{Lyth}.

Once the large scale condition, \eq{lscond}, is violated then $\Phi$
and hence $\R$ will become oscillatory.  In order to quantify whether
$\R$ varies before this , we look at whether there is a significant
relative change in $\R$ in one Hubble time which is approximately
given by the time when
\begin{equation}
\left| \frac{1}{H} \frac{\dot{\R}}{\R} \right| \ge 1. 
\slabel{Rt}
\end{equation}
Assuming the $C_4$ mode is dominant in \eq{Phisr1} and substituting
\eqsss{sr}{Phisr1}{Rconst}{Rdot} into \eq{Rt} gives the condition for significant
variation for $\R$ to be
\begin{equation}
4k^2|\R|\exp(-2|t|/a_0) \ge |\R|
\end{equation}
which implies there will be significant variation in $\R$ due to the
comoving pressure perturbation even on large scales for 
\begin{equation}
\frac{|t|}{a_0} \le \ln(2k).
\slabel{Rvarytime1}
\end{equation}
However, comparing this with the large scale condition, \eq{lsscond1}
and remembering that we have assumed $|\phi_0| \gg \Mp$, it is clear
that the large scale condition is always violated before the condition
in \eq{Rvarytime1}. So, we can conclude that if the $C_4$ mode is
dominant in \eq{Phisr1}, then the large scale condition will
be violated before $\R$ is effected by the comoving pressure
perturbation.

Assuming the $C_3$ mode is dominant in \eq{Phisr1} and assuming
$|\phi_0|/\Mp \gg |t|/a_0$,  substituting \eqs{sr}{Phisr1}
into \eq{Rdot} gives
\begin{equation}
 \frac{1}{H} \dot{\R} 
= -2\Phi_* k^2 \frac{\phi_0^2}{\Mp^2} \exp\left( \frac{|t_*|}{a_0} - 3\frac{|t|}{a_0} \right)
\slabel{Rdot1}
\end{equation}
where $\Phi_* = \Phi(t_*)$ and $t_*$ is arbitrary except for
satisfying the large scale condition \eq{lscond}. Substituting
\eqs{Rdot1}{Rconst} into \eq{Rt} gives the time period for which $\R$ will not
vary on large scales due to the comoving pressure  perturbation
\begin{equation}
\frac{|t|}{a_0} \gtrsim \frac{1}{3} \left[ \ln\left(2 k^2 \frac{\phi_0^2}{\Mp^2} \left|
\frac{\Phi_*}{C_4} \right| \right)+ \frac{|t_*|}{a_0} \right].
\slabel{Rchange}
\end{equation}
If $t_*$ is taken to be the time when the large scale approximation
breaks down, \eq{lsscond1}, then using \eqss{Phisr1}{Rconst}{Rchange}
implies the condition for the comoving pressure perturbation to cause
$\R$ to vary before the large scale condition, \eq{lsscond1} is
violated is
\begin{equation}
\frac{|\Phi_3|}{|\Phi_4|} > \frac{1}{4}
\left(\frac{|\phi_0|}{\Mp}\right)^2 k^2
\end{equation}
where $\Phi_3$ and $\Phi_4$ are the $C_3$ and $C_4$ modes in
\eq{Phisr1}.  It follows that provided the $\Phi_3$ mode is
sufficiently larger than the $\Phi_4$ mode, $\R$ can vary
significantly even while the large scale condition, \eq{lsscond1},
holds. However, for large enough $k$ the modes will always violate the
large scale condition before the comoving pressure effects $\R$.



\eq{Rchange} can be tested against the numerical results in Figures
\ref{fig:Phi} and \ref{fig:Rchange}. Separate values of $\Phi_*$ and
$C_4$ are needed for before and after the bounce. The values of
$\Phi_*$ can be read off for some $|t_*|$ of Figure
\ref{fig:Phi} which satisfies the large scale condition,
\eq{lsscond1} and is where the growing mode is dominating. The values of
$C_4$ can be evaluated from the $\R$ constant parts of Figure
\ref{fig:Rchange}. These values can be substituted into \eq{Rchange}
to determine the region in which the entropy perturbation causes $R$
to vary. These limits are drawn in with dashed vertical lines in
Figure \ref{fig:Rchange}. As can be seen they accurately delimit the
region in which $\R$ changes. From \eq{lsscond1}, it can be seen that
the large scale condition holds for $|t|/a_0 \gtrsim 5$ and so $\R$ is
varying in the collapsing phase even when the large scale condition
holds.


\begin{figure}
\begin{center}
\includegraphics[width=8cm]{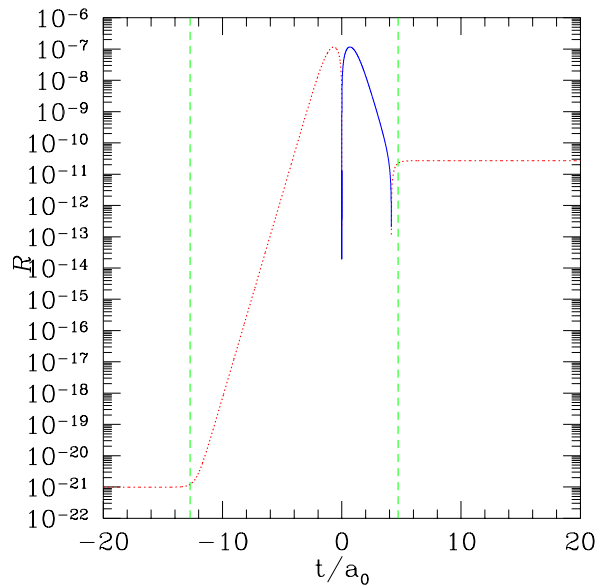}
\end{center}
\caption{\slabel{fig:Rchange} A plot of the absolute value of the
comoving curvature perturbation ($\R$) corresponding to the zero shear
curvature perturbation ($\Phi$) in Figure \ref{fig:Phi} which was for
$n=3$. The solid line is positive values of $\R$ and the dotted line
is negative values. The vertical dashed lines enclose the time domain
in which $\R$ is expected to vary according to \eq{Rchange}.  }
\end{figure}

\eq{Rconst} can be used to write \eq{Phisr} as 
\begin{equation}
\Phi(|t| > |t_\dagger|) \approx C_3 \frac{H}{a}
+ \R \left[1 - \sqrt{\pi} \psi \erfcx(\psi)  \right].
\slabel{Phisr2}
\end{equation}
Where $|t_\dagger|$ satisfies both \eqs{lsscond1}{Rchange}, i.e.\ it
delimits the time on which $\R$ can vary.
The even and odd modes obtained from the numerical solution
in Section \ref{numerics} can be decomposed into growing and decaying 
parts as in \eqss{Phisr}{C3}{C4}
\begin{equation}
\Phi_{\rm even} \approx \left\{
\begin{array}{ll}
C_{\rm even, grow}^- H/a
+ \R_{\rm even}(-|t_\dagger|) \left( 1 - \sqrt{\pi} \psi \erfcx(\psi)  \right),
& t < -|t_\dagger| \\
C_{\rm even, decay}^+ H/a
+ \R_{\rm even}(|t_\dagger|) \left( 1 - \sqrt{\pi} \psi \erfcx(\psi)  \right),
& t > |t_\dagger|
\end{array}
\right.
\slabel{Phisreven}
\end{equation}
A similar equation holds for the odd mode
\begin{equation}
\Phi_{\rm odd} \approx \left\{
\begin{array}{ll}
C_{\rm odd, grow}^- H/a
+ \R_{\rm odd}(-|t_\dagger|) \left( 1 - \sqrt{\pi} \psi \erfcx(\psi)  \right),
& t < -|t_\dagger| \\
C_{\rm odd, decay}^+ H/a
+ \R_{\rm odd}(|t_\dagger|) \left( 1 - \sqrt{\pi} \psi \erfcx(\psi)  \right),
& t > |t_\dagger|
\end{array}
\right.
\slabel{Phisrodd}
\end{equation}
The growing and decaying mode also have even and odd properties:
\begin{equation}
\left. \frac{H}{a} \right|_t = -\left. \frac{H}{a}  \right|_{-t}
\slabel{Hovera}
\end{equation}
and
\begin{equation}
\left.(1 - \sqrt{\pi} \psi \erfcx(\psi)   )\right|_t = \left.(1 -
\sqrt{\pi} \psi \erfcx(\psi)   )\right|_{-t}.
\slabel{constmode}
\end{equation}
The fact that the growing and decaying mode have even and odd
properties does not imply they do not both contribute to $\Phieven$ and
$\Phiodd$. It does mean that it is possible to reduce the number of
unknown coefficients used to describe the even and odd modes. From
\eq{Phisreven}
\begin{equation}
C_{\rm even, grow}^- \left. \frac{H}{a} \right|_{-t}
+ \R_{\rm even}(-|t_\dagger|) \left.( 1 - \sqrt{\pi} \psi \erfcx(\psi)
)\right|_{-t} = 
C_{\rm even, decay}^+ \left. \frac{H}{a} \right|_t
+ \R_{\rm even}(|t_\dagger|) \left.( 1 - \sqrt{\pi} \psi \erfcx(\psi)
)\right|_t
\slabel{Phisreven1}
\end{equation}
from which it follows that
\begin{equation}
C_{\rm even, grow}^- \left. \frac{H}{a} \right|_{-t} = C_{\rm even, decay}^+ \left. \frac{H}{a} \right|_t
\end{equation}
which when combined with \eq{Hovera} implies
\begin{equation}
C_{\rm even, grow}^- = -C_{\rm even, decay}^+.
\end{equation}
Also from \eq{Phisreven1} it follows that 
\begin{equation}
\R_{\rm even}(-|t_\dagger|) \left.( 1 - \sqrt{\pi} \psi \erfcx(\psi)
)\right|_{-t} = 
\R_{\rm even}(|t_\dagger|) \left.( 1 - \sqrt{\pi} \psi \erfcx(\psi)
)\right|_t
\end{equation}
which when combined with \eq{constmode} implies
\begin{equation}
\R_{\rm even}(-|t_\dagger|) = \R_{\rm even}(|t_\dagger|).
\slabel{Reven}
\end{equation}
Similarly, using \eqss{Phisrodd}{Hovera}{constmode} it follows that
\begin{equation}
C_{\rm odd, grow}^- = C_{\rm odd, decay}^+.
\end{equation}
and
\begin{equation}
\R_{\rm odd}(-|t_\dagger|) = -\R_{\rm odd}(|t_\dagger|).
\slabel{Rodd}
\end{equation}
An initially pure growing mode in the collapsing phase could be
constructed as follows
\begin{equation}
\Phi_{\rm grow} = \frac{\Phi_{\rm odd}}{\R_{\rm odd}(-|t_\dagger|)} -
\frac{\Phi_{\rm even}}{\R_{\rm even}(-|t_\dagger|)} 
\slabel{Phigrow}
\end{equation}
provided $\R_{\rm odd}(-|t_\dagger|)$ and $\R_{\rm
even}(-|t_\dagger|)$ are non-zero.
Using \eqsss{Phisreven}{Phisrodd}{Reven}{Rodd}, \eq{Phigrow} becomes
\begin{equation}
\Phi_{\rm grow} = \left\{
\begin{array}{ll}
\frac{H}{a} 
\left( 
\frac{C_{\rm odd, grow}^-}{\R_{\rm odd}(-|t_\dagger|)} -
\frac{C_{\rm even, grow}^-}{\R_{\rm even}(-|t_\dagger|)} 
\right), & t<  -|t_\dagger| \\
\frac{H}{a} 
\left( 
\frac{C_{\rm odd, grow}^-}{\R_{\rm odd}(-|t_\dagger|)} +
\frac{C_{\rm even, grow}^-}{\R_{\rm even}(-|t_\dagger|)} 
\right) - 2 ( 1 - \sqrt{\pi} \psi \erfcx(\psi)
),
& t> |t_\dagger| 
\end{array}
\right.
\slabel{Phigrow1}
\end{equation}
Substituting \eq{Phigrow} into \eq{R} gives
\begin{equation}
\R_{\rm grow} = 
\frac{\R_{\rm odd}}{\R_{\rm odd}(-|t_\dagger|)} - 
\frac{\R_{\rm even}}{\R_{\rm even}(-|t_\dagger|)}
\slabel{Rgrow}
\end{equation} 
and substituting \eq{Phigrow1} into \eq{R} gives
\begin{equation}
\R_{\rm grow} = \left\{
\begin{array}{ll}
0, & t < -|t_\dagger| \\
-2, & t > |t_\dagger|.
\end{array}
\right.
\slabel{result}
\end{equation}
As we are free to multiply \eq{Phigrow} by a $k$ dependent constant,
it follows that $\R_{\rm grow}(t > |t_\dagger|)$ may be some other value
besides minus two. But it is only zero if $\Phi_{\rm grow}=0$, i.e.\
no perturbation at all.  Therefore we have shown analytically that an
ingoing pure growing mode leads to a non-zero outgoing growing mode
provided both the odd and even modes have a $C_4$ component.
However, we can not say what the scale dependence will be without
knowing the scale dependence $\Phi_{\rm grow}$. Figure \ref{fig:R}
shows a plot of the various quantities in \eq{Rgrow}, where
$-|t_\dagger|$ is taken as the minimum time in the figure.
\begin{figure}
\begin{center}
\includegraphics[width=8cm]{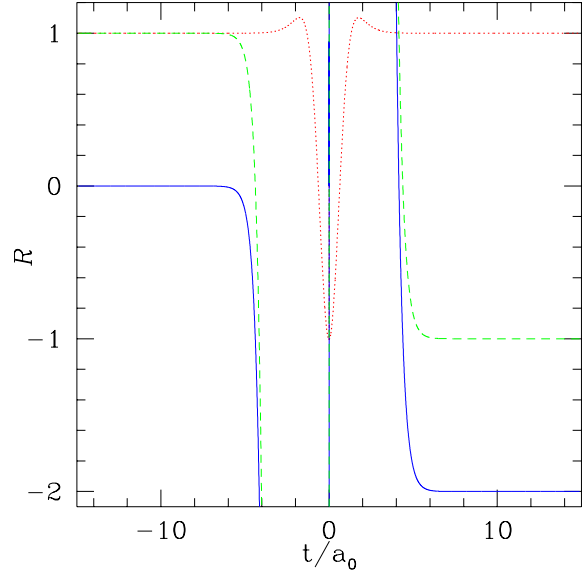}
\end{center}
\caption{\slabel{fig:R} A plot of the comoving curvature perturbation
$\R$ for $n=3$. The initially pure growing mode as constructed in
\eq{Rgrow} is the solid line. The normalised (as in \eq{Rgrow}) even
(dotted line) and odd (dashed line) are also displayed.}
\end{figure}
This numerical integration is in agreement with \eq{result}.

\section{Conclusion}
In this article we have examined the way perturbations evolve across a
bounce in a scalar field dominated closed Universe. The background
model naturally bounces without violating General Relativity or the
Weak Energy Condition. Analytical approximations were found for the
background and perturbations. The perturbation approximations were
found to work well around the bounce $|t|\lesssim a_0$ and away from
the bounce in the slow-roll regime $|t|\gtrsim a_0$. However the
overlap between the approximations was not sufficient to use the
analytical approximations alone for arbitrary initial conditions. But
they were useful in checking and interpreting the analytical
solutions. The numerical solutions were complicated by the ordinary
differential equation for $\Phi$ having a singular coefficient at the
bounce. Thus a Taylor approximation had to be used around the bounce
and shooting methods used to obtain the numerical solution for
particular initial conditions for a wider domain. The Taylor expansion
around the bounce could also be used to show that the perturbations
and their first and second time derivatives are finite at the bounce
and so non-linearities in the Einstein tensor are not inevitable, but
depend on the choice of initial conditions.

It was shown that all wave number modes violate the large scale
condition for $|t|\gtrsim a_0$ and so become oscillatory around the
bounce. We also found during the collapsing phase the growing mode
could cause the comoving pressure to increase sufficiently to make
$\R$ vary significantly as the bounce is approached. The same is true
for the decaying mode as the bounce is approached from the expanding
phase side.  Thus, in general we do not expect $\R$ to be constant
around the bounce. A formula predicting when $\R$ starts to vary was
given.  We also showed numerically that, for this model, a pure
ingoing growing mode ($\R^-=0$) did not lead to a pure outgoing
decaying mode ($\R^+=0$).
	
This model shows that in general the dynamics of the bounce need to be
taken into account when evaluating how perturbations change across a
bounce.

{\bf Acknowledgements} We would like to thank M. Bucher, D. Lyth and D. Wands for
helpful discussions. This research is funded by PPARC (UK). 

\bibliographystyle{plain}

\end{document}